\documentclass[twocolumn,secnumarabic,amssymb, nobibnotes, pra,superscriptaddress]{revtex4-2}
\usepackage{CJK}
\usepackage{graphicx}
\usepackage{blindtext}
\usepackage{physics}
\usepackage{siunitx}
\usepackage{lipsum}
\usepackage{amsfonts,amssymb,amsmath}
\usepackage{bm}
\usepackage{xcolor,calc}
\usepackage{ulem}
\usepackage{hyperref}
\usepackage{tikz}
\usepackage{quantikz}
\usepackage{float}
\usepackage{hyperref}
\usepackage{bbold}
\newcommand{\gev}{GeV }
\newcommand{\gevv}{GeV}

\newcommand{\nuc}{${}^{13}$C }
\newcommand{\nucc}{${}^{13}$C}
\newcommand{\uproman}[1]{\uppercase\expandafter{\romannumeral#1}}

\newcommand{\fomav}{$\text{FoM}_{\text{av}}$}
\newcommand{\fomcb}{$\text{FoM}_{\text{cb}}$}
\newcommand{\Fav}{\mathcal{F}_{\text{av}}}

\newcommand{\id}{\mathbb{1}}
\begin{document}
\title{Optimal Two-Qubit Gates for Group-IV Color-Centers in Diamond} 
\author{Jurek Frey}
\thanks{Corresponding author: \href{mailto:ju.frey@fz-juelich.de}{ju.frey@fz-juelich.de}}
\affiliation{Peter Gr\"unberg Institute-Quantum Computing Analytics (PGI-12), Forschungszentrum J\"ulich GmbH, D-52425 J\"ulich, Germany}
\affiliation{Theoretical Physics, Saarland University, D-66123 Saarbrücken, Germany}
\author{Katharina Senkalla}
\affiliation{Institute for Quantum Optics, Ulm University, Albert-Einstein-Allee 11, D-89081 Ulm, Germany}
\author{Philipp J. Vetter}
\affiliation{Institute for Quantum Optics, Ulm University, Albert-Einstein-Allee 11, D-89081 Ulm, Germany}
\author{Fedor Jelezko}
\affiliation{Institute for Quantum Optics, Ulm University, Albert-Einstein-Allee 11, D-89081 Ulm, Germany}
\author{Frank K. Wilhelm}
\affiliation{Peter Gr\"unberg Institute-Quantum Computing Analytics (PGI-12), Forschungszentrum J\"ulich GmbH, D-52425 J\"ulich, Germany}
\affiliation{Theoretical Physics, Saarland University, D-66123 Saarbrücken, Germany}
\author{Matthias M. M\"uller}
\affiliation{Peter Gr\"unberg Institute-Quantum Control (PGI-8), Forschungszentrum J\"ulich GmbH, D-52425 J\"ulich, Germany}
\date{\today}
\begin{abstract}
Color centers associated with group-IV dopants in diamond with long-lived nuclear spins have emerged as major candidates for distributed quantum computing nodes and quantum repeaters. Several proof-of-principle experiments have already been demonstrated. A key operation for long-distance entanglement-distribution protocols are fast and robust gates between the electron spin and a nuclear spin.
Here, we investigate numerically for an existing experimental platform of a Germanium-vacancy (GeV) center with a strongly-coupled \nuc spin, how such gates can be implemented via quantum optimal control. In the presence of realistic noise we investigate different parameter regimes and gate operations and obtain robust two-qubit gates with fidelities exceeding $99.9 \%$.
The framework provides a scalable strategy for group-IV quantum nodes and can be adapted to related architectures.
\end{abstract}
\maketitle
\section{Introduction}
Group-IV color centers in diamond have shown tremendous potential as platforms for quantum-network nodes, capable of distributing entanglement over long distances~\cite{knaut2024entanglement, ruf2021quantum}, which is crucial for applications including distributed quantum computing~\cite{wei2025universal} and non-local sensing~\cite{stas2025entanglement}.
These color centers exhibit strong optical transitions with transform limited spectral lines and long ground state spin coherence times. 
Surrounding \nuc nuclear spins can serve as long-living memory qubits~\cite{metsch2019initialization,grimm2025coherent,wei2025universal,Klotz2026,resch2025high} or can be used as a resource for error correction as has been demonstrated for nitrogen-vacancy (NV) centers in diamond~\cite{taminiau2014universal,childress2006coherent, bradley2019ten}.
Note that for group‑IV color centers the control of nuclear spins is generally more challenging compared to NV centers due to the unavailability of intrinsic host nuclear spins~\cite{harris2023hyperfine} and the symmetry of the hyperfine interaction due to the spin-1/2 structure~\cite{beukers2025control}.

A color center is typically selected based on suitable optical and coherence properties such as high readout fidelity, long coherence times, or favorable coupling to an optical cavity.
Although approaches exist to engineer a deterministic nuclear-spin environment, such as incorporating \nucc-enriched layers near the color centers~\cite{vetter2025room}, the precise positions of individual \nuc spins remain inherently random.
The detailed configuration of nearby nuclear spins is often determined at a later stage and is not usually a primary factor in the initial selection ~\cite{robledo2011high, bradley2019ten, grimm2025coherent}.
However, a \nuc nuclear spin may only be a few lattice sites away, resulting in strong hyperfine interaction.
While this enables fast two-qubit gate operations, it can also degrade gate fidelities if not treated carefully ~\cite{Gundlapalli2025, resch2025high, beukers2025control,Klotz2026}. \\
Here, we demonstrate that quantum optimal control (QOC)~\cite{glaser2015training,muller2022one,rembold2020, koch2025} can overcome the challenges posed by strong electron–nuclear hyperfine coupling, enabling robust two-qubit gates with high fidelity.
In particular, we employ the quantum optimal-control suite QuOCS~\cite{rossignolo2023quocs} with its dCRAB algorithm~\cite{rach2015dressing, muller2022one} to optimize the pulse shapes in an open-loop manner, while modeling decoherence through an Ornstein–Uhlenbeck (OU) process~\cite{uhlenbeck1930theory, grimm2025coherent}.
We base our analysis on a well-studied experimental setup of a germanium-vacancy (\gevv) center coupled to a nearby \nuc nuclear spin~\cite{senkalla2024germanium, grimm2025coherent}, where the hyperfine interaction has so far imposed an inherent limit on achievable gate fidelities~\cite{Gundlapalli2025}. \\
We model the dynamics across a wide range of magnetic-field settings and identify distinct operating regimes that support high-fidelity control.
We achieve gate fidelities of $99.91\%$ and $99.94\%$ for SWAP and CNOT gates, respectively.
For the SWAP gate we also employ QOC with algebraic gate decomposition to optimize only the non-local gate component~\cite{Mueller2011,watts2015optimizing,Goerz2015,muller2022one}.
With this additional degree of freedom we decrease the gate error by almost one order of magnitude.
Since group-IV color centers host an electron spin-1/2 system, our results are relevant across this class of defects ~\cite{ruf2021quantum, bradac2019quantum}.
The paper is now structured as follows.
We first introduce the system. We then discuss the requirements for a gate set for a quantum node. Then, we proceed to presenting our QOC methods and the results.
\section{System}
\label{sec:system}
We consider a two-qubit system consisting of the electron spin of the negatively-charged GeV (given by the $S_k$ spin operators; $k=x,y,z$) and a strongly-coupled nuclear spin (given by the $I_k$ spin operators; $k=x,y,z$), i.e., the coupling is stronger than environmental decoherence.
Setting $\hbar=1$, we obtain the Hamilton operator in the secular approximation with respect to the electronic Zeeman splitting:
\begin{equation}
    H_0 = -\omega_e S_z - \omega_I I_z + A_{zz} S_z I_z + A_{zx} S_z I_x, 
\end{equation}
with the electronic $\omega_e = \gamma_e B$ and nuclear $\omega_I = \gamma_I B$ Zeeman splittings at the magnetic field $B$.
Here,  $\gamma_e, \gamma_I$ are the effective gyromagnetic ratios of the electronic and nuclear spin.
The interaction between the electron and the nuclear spin is described by the longitudinal and transverse hyperfine couplings, $A_{zz}$, and $A_{zx}$, respectively. 
For the hyperfine couplings, we choose $A_{zz} = 2\pi \times \SI{2.86234}{\MHz}$, and $A_{zx} = 2\pi \times \SI{0.60281}{\MHz}$, according to Ref.~\cite{grimm2025coherent}. 
These hyperfine parameters are intrinsic and depend only on the position of the nuclear spin relative to the defect.
However, the effective projections depend on the magnetic field direction and strength.
The latter one presents an easily tunable parameter reflected by the Larmor frequency $\omega_I$, which we explore in the few MHz range. 
The precise values will be specified in the respective sections. 
\\
The dynamics in the system are coherently controlled via time-dependent driving of the electronic transitions in the GeV center. 
The nuclear spin is addressed indirectly through the hyperfine couplings, since direct nuclear control suffers from a slow Rabi frequency, while the strong electron-nuclear coupling allows for faster implementations ~\cite{hegde2020efficient, vallabhapurapu2022indirect,frey2026}. 
The control Hamiltonian $H_c$ (in the electron frame) takes the form 
\begin{align}
    H_c(t) = \Omega_{\text{MW}}(t)
  \!\left[\cos(t \Delta - \phi)\,S_x
   -\sin(t \Delta - \phi)\,S_y\right]
\end{align}
where $\Omega_{\text{MW}}$ is the transverse microwave control amplitude in the frame of the electronic Zeemann term $\omega_e$ after the rotating wave approximation, $\phi$ is the phase of the control and $\Delta$ is the detuning of the drive frequency from the electron frequency $\omega_e$.
The experimental limits are incorporated by ensuring $\abs{\Omega_{\text{MW}}} \leq \SI{15}{\MHz}$ \cite{grimm2025coherent}. 
\\
In practical implementations, the fidelity of multi-qubit gates is not only limited by the coupling strengths and imperfections of the control and supply hardware, but also by decoherence of the electron spin \cite{Gundlapalli2025, dobrovitski2013quantum, dolde2014high}. 
To account for this, we extend our system model to include dephasing noise on the GeV electron, modeled as an Ornstein–Uhlenbeck (OU) process $\beta(t)$ with parameters extracted from experimental measurements \cite{grimm2025coherent}.
We incorporate a semi-classical dephasing term in the Hamiltonian
\begin{align}
    H_{\text{noise}}(t) = \beta_k(t) S_z, 
\end{align}
where the stationary, Gaussian noise process $\beta(t)$ is generated by discretizing the corresponding differential equation and sampled over different realizations. 
Summarizing, the total Hamiltonian in the electron frame of the system 
\begin{align}
    H(t) = H_0 +\omega_e S_z + H_c(t) + H_{\text{noise}}(t).
\end{align}
fully specifies the system dynamics considered in this work.
For more details we refer to the appendix.
\section{Quantum Node Requirements}
\subsection{Quantum Circuits on the Node}
At the circuit level, the essential operations on a single node required for remote entanglement distribution and quantum error correction can be expressed by a small set of fundamental one- and two-qubit primitives, most notably the Hadamard, CNOT, and SWAP gates ~\cite{nickerson2014freely}.
Combined with single-shot readout and photon-mediated electron-electron entanglement, these primitives form the building blocks for state preparation, remote entanglement generation, swapping, and purification protocols ~\cite{dur1999quantum,azuma2023quantum}. 
Practical implementations are inevitably subject to noise and experimental imperfections, such that only gate operations that are both robust and fast can support scalable quantum-network protocols with acceptable success rates and memory lifetimes~\cite{nickerson2014freely, fowler2012surface}.
\\
Within the physical constraints, it is important to find hardware-efficient implementations of these primitives~\cite{dobrovitski2013quantum}.
In the considered \gev platform with a strongly-coupled, long-lived \nuc \cite{grimm2025coherent}, these are realized in a native way using fast electron–nuclear interactions~\cite{grimm2025coherent, Gundlapalli2025}.
Similar approaches have enabled remote entanglement and quantum network protocols in related platforms \cite{beukers2025control,kalb17entanglement,pompili21realization,wei2025universal,stolk24metropolitan}.
In our setting, the found logical Hadamard, CNOT, and SWAP operations are locally equivalent to the ideal circuit primitives up to known single-qubit phases, which can be compensated by calibrated control pulses, tracked via Pauli-frame updates or virtual gates~\cite{makhlin2002nonlocal, knill2005quantum}.
These would allow for measurement-based state preparation of the nuclear memory, which on a logical level correspond to a Hadamard gate on the electron spin followed by a CNOT-equivalent and projective measurement~\cite{Gundlapalli2025,robledo2011high, taminiau2014universal}.
Storage and reuse of the memory are then enabled by SWAP- or equivalent gates that coherently transfer quantum states and entanglement between the electron spin and the nuclear spin~\cite{dolde2014high, waldherr2014quantum}.
\subsection{Two-Qubit Gates}
\label{sec:two-qubit}
In particular, we want to show how to realize a two-qubit gate set consisting of an electron-controlled NOT on the nuclear target  C$_e$NOT$_n$ and its nuclear-controlled counterpart, the C$_n$NOT$_e$ gate, as well as a SWAP gate.
The full SWAP operation is not strictly required for all protocols but is useful for state transfer and therefore storing information on the nuclear spin state.
We can further analyze the two-qubit gates in terms of local symmetries. E.g., one can obtain the C$_e$NOT$_n$ by applying the C$_n$NOT$_e$ gate together with a Hadamard gate on each qubit before and after the CNOT. 
More in general, one can consider the
Cartan (or KAK) decomposition~\cite{Khaneja2001cartan, Zhang2003b,Zhang2004a,Nakajima2005ANA}
(see also~\ref{sec:remarks}) of a two-qubit gate $U$,
\begin{equation}
U = K_1 A K_2,
\end{equation}
where $K_1$ and $K_2$ are local gates and the nonlocal component $A$ takes the canonical form
\begin{equation}
A = \exp\!\left[-\frac{i}{2}
\left(c_1\, XX + c_2\, YY + c_3\, ZZ\right)\right],
\end{equation}
where $X$, $Y$, and $Z$ are the respective Pauli matrices.
The coefficients $c_i$ are the so-called Weyl-chamber coordinates.
The gate decomposition can also be used to show the universality of a gate set and decompose a given quantum circuit into the available gates~\cite{Zhang2004a, Nakajima2005ANA,wang2025}.
Both CNOT gates share the same Weyl-coordinates ($c_1=\pi/2$, $c_2=c_3=0$) and differ only by local, i.e. single qubit, rotations ~\cite{zhang2003geometric, makhlin2002nonlocal}; in this case the aforementioned Hadamard gates.
The SWAP gate instead can be decomposed into three CNOT gates, alternating the role of control and target between the two qubits.
The SWAP gate has the Weyl chamber coordinates $c_1=c_2=c_3=\pi/2$ and therefore is the maximally non-local operation in the two-qubit Weyl chamber ~\cite{zhang2003geometric}.
This is also why its direct generation (from a single control pulse) provides a good benchmark for control performance.
By using QOC, we will design pulses for all three two-qubit gates as well as for an arbitrary gate in the local equivalence class of SWAP.

\begin{figure}
\centering
\includegraphics{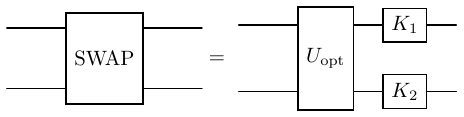}
\caption{Circuit diagram of a SWAP alternative $U_\text{opt}$. If the non-local content of the optimized gate corresponds to the SWAP gate, i.e. $\mathcal{F}_{\text{nl}}=1$ (see Eq.~\eqref{eq:fnl}), the two gates differ just by a local transformation on each of the qubits.}
\label{fig:SWAP_decomp}
\end{figure}

To illustrate the decomposition, in \autoref{fig:SWAP_decomp} we show a circuit-level sketch of the SWAP decomposition which clarifies, how the SWAP gate relates to the optimized gate $U_{\text{opt}}$.
The local operations that transform $U_{\text{opt}}$ into the SWAP gate can be obtained by
\begin{equation}
    K_1 \otimes K_2 = U_{\text{opt}}^\dagger \text{SWAP}.
\end{equation}
The formula is derived in \autoref{sec:remarks}.
Alternatively, one could also apply the local gates just on one qubit, e.g. by $\text{SWAP}=(K_2\otimes\id) U_{\text{opt}} (K_1 \otimes\id)$. 
Within a larger quantum circuit the local gates can often be combined with other local gates or performed only virtually, so that they do not necessarily cause an additional overhead.
\section{Quantum Optimal Control}
\label{sec:optimal_control}
The challenges due to dephasing and strong hyperfine interaction have motivated the use of advanced control techniques, including shaped microwave pulses, dynamical decoupling, and quantum optimal control, which have been shown to significantly improve gate fidelities in solid-state spin registers in diamond~\cite{Said2009,dolde2014high, dobrovitski2013quantum,scheuer2014precise,Frank2017,rembold2020,Chen2020,Oshnik2022,rossignolo2023quocs,Takou2023,vetter2024gate,wang2025,lim2025,baran2026,marcomini2026}.
Here, for the optimization of the two-qubit gates, we consider two distinct optimal-control approaches.
Both take into account the same system, control and noise model which was introduced in \autoref{sec:system}.
In the first approach, our goal is to generate a unitary $V_{G}$, where $V_G \in \{ \text{ C$_e$NOT$_n$,  C$_n$NOT$_e$, SWAP}\}$.
For each sampled noise realization the application of a control pulse induces a time-evolution which yields a final unitary $U_k(T)$.
The fidelity averaged over $N$ noise realizations can be defined as:
\begin{align}
        \mathcal{F}_{\text{av}}(V_G) &= \frac{1}{N} \sum_{k=1}^N   \frac{1}{d^2}\abs{\Tr{U_k^\dagger(T) V_G}}^2.
\label{eq:multi_F_aver}
\end{align}
The figure of merit (FoM) is then:
\begin{align}
        \text{FoM}_{\text{av}}(V_G) &= 1-\mathcal{F}_{\text{av}}(V_G).
\label{eq:multi_FoM_aver}
\end{align}
Here, $d=4$ is the dimension of the two-qubit Hilbert space, $T$ is the gate duration and minimizing this FoM corresponds to a minimization of the average distance between the realized operation and the target gate $V_G$.\\\\
For many applications, only the nonlocal part of a two-qubit gate is relevant, since gates that differ solely by local single-qubit operations are equivalent with respect to their entangling power and computational capability ~\cite{makhlin2002nonlocal, kraus2001optimal}. 
This motivates an approach in which we focus on optimizing for gates that are non-locally equivalent, rather than targeting a specific representative within a given equivalence class as has been demonstrated for various platforms~\cite{Zeier2007,Mueller2011, watts2015optimizing,Goerz2015, Yuan2015, Goerz2017,muller2022one}, including for NV center spin registers~\cite{wang2025,baran2026}.
Here, we follow the formulation of Ref.~\cite{watts2015optimizing} with a slight adaption of the FoM that allows us to optimize for pulses that are robust against the dephasing noise.

Every target operation $V_G$ is uniquely (up to local unitaries) characterized by its Weyl coordinates $\{c_1^{(G)},c_2^{(G)},c_3^{(G)}\}$. 
For every controlled time evolution gate we calculate the non-local invariants obtained from the time-evolution according to the control pulse without any noise $U_{\text{ideal}}$ (ideal time evolution).
The invariants obtained from $U_{\text{ideal}}$ can be used for introducing a measure of how close the gate corresponds to a locally equivalent target gate.
This nonlocal, noiseless measure is given via \cite{watts2015optimizing}
\begin{align}
    \mathcal{F}_{\text{nl}}(V_G, U_{\text{ideal}}) = \cos(\frac{\Delta c_1}{2})\cos(\frac{\Delta c_2}{2})\cos(\frac{\Delta c_3}{2}),
    \label{eq:fnl}
\end{align}
where $\Delta c_i = c_i^{(G)}- c_i$ are the Weyl coordinate differences. 
With this we introduce the combined FoM as 
\begin{equation}
    \text{FoM}_{\text{cb}}(V_G) = 1 - \mathcal{F}_{\text{nl}}(V_G, U_{\text{ideal}}) + \text{FoM}_{\text{av}}(U_{\text{ideal}}).
    \label{eq:combined_FoM}
\end{equation}
Here, the first term enforces the correct nonlocal gate content in the absence of noise and the second term penalizes its sensitivity to noise by calculating the overlap of the noiseless gate $U_{\text{ideal}}$ with the noisy realizations $U_k(T)$.
This second approach with the combined $\text{FoM}_{\text{cb}}$ allows us to find a suitable representative of the target equivalence class, prioritizes the robustness of the nonlocal interactions and is suited to be embedded in larger circuits.
The details of the optimization parameters can be found in \autoref{sec:num_methods}.
\section{Results}
\subsection{Robust SWAP and CNOT optimization under realistic noise}
\label{sec:robust}
\begin{figure}
    \centering
    \includegraphics{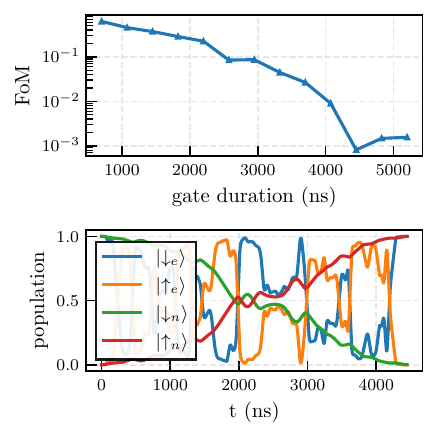}
        \caption[Optimization process of a C$_\text{e}$NOT$_\text{n}$ gate]{Optimization process of a $V_G=$ C$_\text{e}$NOT$_\text{n}$ gate the upper panel shows the reached FoM for different gate durations, the best reached fidelity was  $\mathcal{F}_{\text{av}}(V_G) = \SI{99.92 \pm 0.005} \%$ at a gate duration of $\SI{4.450}{\micro \s}$ and averaged over $N=5000$ noise realizations, the mean and error here was obtained by evaluating the pulse 100 times. The lower panel shows the populations resulting from the optimized pulse with a gate duration of $\SI{4.450}{\micro \s}$ while the other spin is traced out.}
    \label{fig:CNOT_N}
\end{figure}

\begin{figure}
    \centering
    \includegraphics{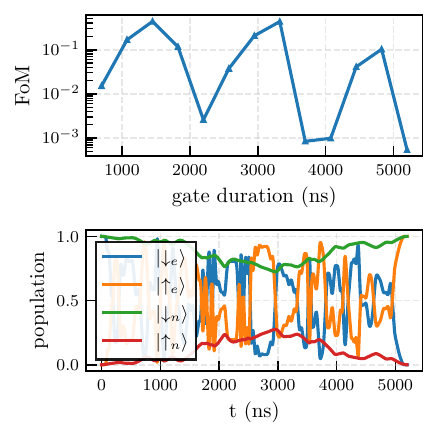}
    \caption[Optimization process of a C$_\text{n}$NOT$_\text{e}$ gate]{Optimization process of a $V_G =$  C$_\text{n}$NOT$_\text{e}$ gate the upper panel shows the reached FoM for different gate durations, the best reached fidelity was $\Fav(V_G) =\SI{99.942 \pm 0.003}\%$ at a gate duration of $\SI{5.2}{\micro \s}$ and averaged over $N=5000$ noise realizations. The lower panel shows the populations resulting from the optimized pulse with gate duration of ($\SI{5.2}{\micro \s}$) while the other spin is traced out.}
    \label{fig:CNOT_E}
\end{figure}

\begin{figure}
    \centering
    \includegraphics{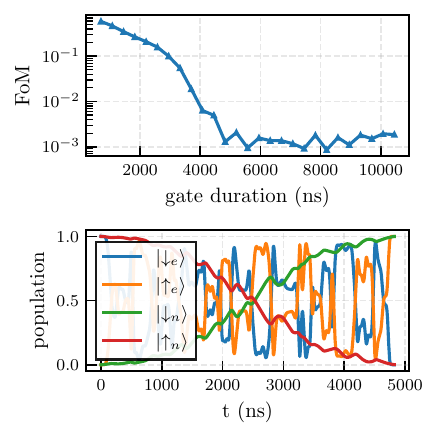}
    \caption{Optimization process of a $V_G =$ SWAP gate the upper panel shows the reached FoM for different gate durations, the best reached fidelity was  $\Fav(V_G) =\SI{99.914 \pm 0.007} \%$ $(\SI{99.874 \pm 0.006} \%)$ at a gate duration of $\SI{8.2}{\micro \s}$ ($\SI{4.825}{\micro \s}$) and averaged over $N=5000$ noise realizations. The lower panel shows the populations resulting from the optimized pulse with gate duration of $\SI{4.825}{\micro \s}$ while the other spin is traced out.}
    \label{fig:SWAP}
\end{figure}

High-fidelity multi-qubit gates have to provide robustness against realistic noise, in this case especially against the dominant electron dephasing ~\cite{khodjasteh2009dynamical, dobrovitski2013quantum, dolde2014high,senkalla2024germanium,grimm2025coherent}.
To do this the FoM was chosen according to Eq.~\eqref{eq:multi_FoM_aver}.
We show here results for the electron-controlled C$_{\text{e}}$NOT$_{\text{n}}$ gate (see \autoref{fig:CNOT_N}), the nuclear-controlled C$_{\text{n}}$NOT$_{\text{e}}$ gate (see \autoref{fig:CNOT_E}) and the SWAP gate (see \autoref{fig:SWAP}) between the electronic and nuclear spin.
We choose the Larmor frequency as $\omega_I = \SI{1.04}{\MHz}$ \cite{grimm2025coherent}, a value that enables to implement high Rabi frequencies of up to $15\,$MHz in the experiment.
For each candidate pulse, the FoM Eq.~\eqref{eq:multi_FoM_aver}  was evaluated over $N=5000$ noise realizations, ensuring convergence towards pulses that are robust against slow dephasing, i.e fluctuations that are slow on the timescale of the control pulse. 
In \autoref{fig:CNOT_N} we show the reached pulse performance of the C$_{\text{e}}$NOT$_{\text{n}}$ gate for different gate durations and the corresponding population dynamics reached by tracing out the opposite spin. 
We obtain a maximum fidelity of $\Fav(\text{C$_{\text{e}}$NOT$_{\text{n}}$}) =99.91 \%$ for a gate time of $T = \SI{4.450}{\micro \s}$, which is approximately one order of magnitude shorter compared to direct driving using RF control \cite{grimm2025coherent}.
In \autoref{fig:CNOT_E} we show the same plots for the C$_{\text{n}}$NOT$_{\text{e}}$ gate.
Here, we achieve an even better fidelity  of $\Fav(\text{C$_{\text{n}}$NOT$_{\text{e}}$}) =99.94\%$ at a gate duration of $\SI{5.2}{\micro \s}$. 
The oscillations in the FoM are likely coming from the $A_{zx}$ contributions. 
In \autoref{fig:SWAP} we show the plots for the SWAP gate, where we achieved a fidelity of  $\Fav(\text{SWAP})=99.91 \%$ at a gate duration of $\SI{8.2}{\micro \s}$  and only a slightly worse but fast alternative with a fidelity of $99.89 \%$ and a gate duration of $\SI{4.825}{\micro \s}$. 
For the realization of Hadamard gates (both on the electron and the nuclear spin) we refer to \autoref{sec:Hadarmards}.
These results demonstrate that optimal microwave control can circumvent the time-limitations imposed by perturbative schemes (such as double-quantum-transition gates) and the hardware limitations which are needed to implement gates via dynamical decoupling sequences in strongly coupled systems ~\cite{taminiau2014universal, van2012decoherence, scheuer2014precise,stas22robust,frey2026}. 
Moreover, the method is flexible: different target unitaries, gate times, or robustness criteria (e.g.\ against detuning noise) can be incorporated directly into the optimization cost function, making it a powerful tool for practical two-qubit gate synthesis in group-\uproman{4} vacancy systems.
\subsection{Tuning the Gate Parameters}
\label{sec:discussion}
\begin{figure}
    \centering
    \includegraphics{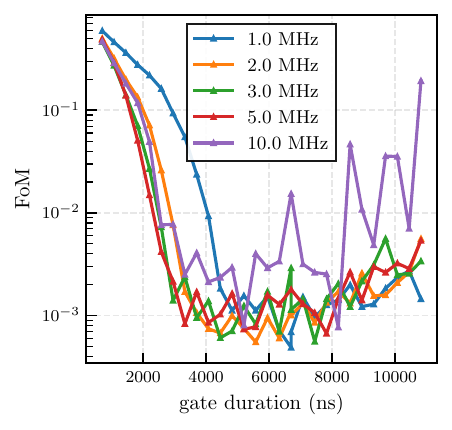}
    \caption{Optimization process of a SWAP gate for different Larmor frequencies and gate durations. The shown curves are optimized with $N_{\text{opt}} = 100$ noise realizations to reduce FoM$_\text{av}$ and are evaluated with $N_\text{eval} = 5000 $ noise realizations. We see that while increasing the Larmor frequency does not necessarily improve the best found FoM, it reduces the gate duration significantly.}
    \label{fig:SWAP_vs_wn}
\end{figure}

\begin{figure}
    \centering
    \includegraphics{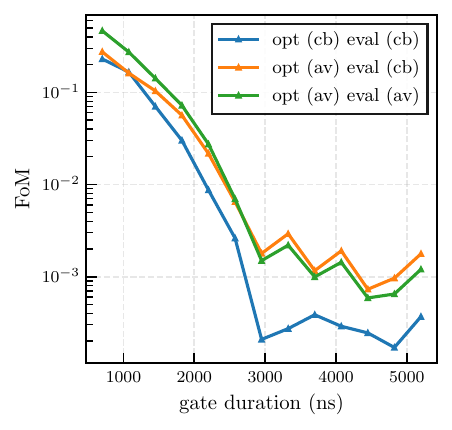}
    \caption{Shown is the performance of a SWAP gate as well as a gate of the same non-local equivalence class for different FoMs: The green curve shows the optimization of the SWAP gate with the standard average fidelity $\text{FoM}_{\text{av}}$ from Eq.~\eqref{eq:multi_FoM_aver}. The orange curve shows the evaluation of the same pulses with Eq.~\eqref{eq:combined_FoM} and the blue curve shows the evaluation of \fomcb\, optimized with \fomcb. We see that we reach \fomcb $= \SI{2.2(1)e-4}{} (\SI{1.8(2)e-4}{})$ for a gate duration of $\SI{2.950}{\micro \s}$ ($\SI{4.825}{\micro \s}$).}
    \label{fig:SWAP_inv}
\end{figure}

In quantum memory applications, the backaction of the readout process on the nuclear spin is a critical consideration, as repeated measurements or imperfect control can lead to unwanted nuclear spin flips or dephasing. 
In solid-state spin registers, single-shot readout of a nuclear spin is typically implemented by mapping the nuclear state onto the electron spin via a CNOT-type gate, followed by optical readout of the electron~\cite{robledo2011high, taminiau2014universal,Gundlapalli2025}.  
While this approach enables high-fidelity and fast readout, the required electron–nuclear entangling gate inevitably exposes the nuclear spin to backaction originating from electron decoherence, control imperfections, and measurement-induced collapse of the electron state~\cite{jiang2009repetitive, robledo2011high}. 
In particular, during the CNOT operation, fluctuations of the electron spin phase or population can be transferred to the nuclear spin through the hyperfine interaction, thereby limiting the robustness of the nuclear quantum memory~\cite{waldherr2014quantum} constraining the single‑shot readout fidelity in the underlying experimental setup~\cite{Gundlapalli2025}.
Here, the CNOT gates optimized in \autoref{sec:robust} can be employed for single-shot readout, while simultaneously minimizing measurement backaction on the nuclear spin by reducing gate duration and suppressing sensitivity to electron dephasing as well as the spin-flip probability induced by the perpendicular hyperfine-coupling term. 
Another approach to additionally increase the gate fidelity is to exploit the fact that the nuclear spin Larmor frequency $\omega_I$ is adjustable via the external magnetic field.
This provides an additional degree of freedom because changing $\omega_I$ modifies the effective coupling strength present in the system.
This effect directly influences the controllability of the electron‑nuclear two‑qubit dynamics and opens up additional regions in parameter space where high‑fidelity gates can be realized. In Fig.~\ref{fig:SWAP_vs_wn} we show the performance of the SWAP gate
versus gate duration for different choices of $\omega_I$.
We observe that higher Larmor frequencies do not lead to a reduction in the achievable fidelity but offer the possibility to increase the achievable gate speed considerably.
However, for very large Larmor frequencies the coherent control of the system seems to become more difficult since we notice an increase in the variance of the FoM. 
We conclude that for future experiments, this should be considered as a resource to enable faster gates.\\
To further improve the fidelity of SWAP-like gates, we now consider the invariant-based optimization. 
In \autoref{fig:SWAP_inv} we compare the standard average-fidelity FoM  (\fomav) optimization with the invariant-based FoM (\fomcb). In particular, the green curve shows the optimization according to \fomav. The orange curve shows the result of the same optimization, but for each value of the gate duration the optimized pulse is evaluated according to \fomcb. Since both curves are very close to each other, we conclude that the two FoMs agree on the evaluation of the gate performance if the gate is close to a SWAP gate.
We then proceed to an optimization according to \fomcb.
We observe that this invariant-based optimization (blue curve) significantly (approximately one order of magnitude) improves the gate performance compared to the orange or green curve. We cannot (meaningfully) evaluate these pulses via \fomav\, because the gates we obtain are not close to a SWAP gate but rather share the same non-local content with the SWAP gate. Instead they differ by local gates as illustrated in Fig.~\ref{fig:SWAP_decomp} and explained in Sec.~\ref{sec:two-qubit}. 
The explicit form of the local gates as well as more details on the decomposition can be found in the appendix \autoref{sec:remarks}. 
We emphasize that the improvement from this invariant-based approach does not simply come from the different choice of the FoM (as can be seen by comparing the blue to the orange curve) but from the fact that the obtained gates correspond more closely to the non-local content of the SWAP gate and are more robust against dephasing.
\section{Conclusion}
\label{sec:conclusion}
Summarizing, we showed that using a dCRAB-based optimal control ansatz, we were able to realize excellent C$_{\text{e}}$NOT$_{\text{n}}$, C$_{\text{n}}$NOT$_{\text{e}}$  and SWAP gates, with fidelities exceeding $99.9\%$ under experimentally motivated noise realizations. 
We demonstrated by varying the nuclear Larmor frequency and the gate duration that we can find regions of the experimental parameter space where high fidelity operations are achievable, providing practical guidance for future experiments. 
Moreover, with Eq.~\eqref{eq:combined_FoM} we employed local symmetries to find new two-qubit gates for this system through the invariant-based SWAP optimization. We have further improved this invariant-based gate optimization for noisy systems through a modified way to introduce robustness.
This allowed us to implement a flexible gate design which relaxes local constraints and offers a new class of robust two-qubit gates suitable for quantum network applications.
\\
The identified control pulses and parameter regimes are directly compatible with current experimental capabilities and therefore allow for experimental verification, e.g., assisted by a suitable closed-loop calibration protocol~\cite{marcomini2026}. 
Overall, our findings support the use of the GeV in quantum networks
and related applications, including fault-tolerant entanglement distribution, distributed quantum computing, and non-local sensing \cite{grimm2025coherent, muller2022one, egger2014adaptive}, which methods are also easily applicable to other group-IV color centers. 
\begin{acknowledgments}
We thank Genko G. Genov, Prithvi Gundlapalli, and Simon G. Walliser for helpful discussions.
This work was supported by Germany’s Excellence Strategy – Cluster of Excellence Matter and Light for Quantum Computing (ML4Q) EXC 2004/1 – 390534769,  the European Union’s HORIZON Europe program via projects SPINUS (No. 101135699) and OpenSuperQPlus100 (No. 101113946), by AIDAS-AI, Data Analytics and Scalable Simulation, which is a Joint Virtual Laboratory gathering the Forschungszentrum J\"ulich and the French Alternative Energies and Atomic Energy Commission, as well as by BMFTR via the projects SPINNING (No. 13N16210 and 13N16215), DE BRILL (No. 13N16207), Quanten4KMU (No. 03ZU1110BB), QuantumHiFi (No. 16KIS1593), QR.N, CoGeQ and Deutsche Forschungsgemeinschaft (DFG) via project No. 387073854 and joint DFG-Japan Science and Technology Agency (JST) project ASPIRE (No. 554644981).
\end{acknowledgments}
\section*{Data Availability}
The data presented in this study is available under \url{https://doi.org/10.5281/zenodo.18832255}.
\section*{Author Contributions}
J.F. performed the numerical simulations and analytical calculations. J.F., K.S., P.J.V. and M.M.M. wrote the manuscript. M.M.M., F.K.W. and F.J. supervised the project. All authors contributed to the discussion of the results.
\appendix
\section{Model and Hamiltonian}
\label{sec:Model_and_Hamiltonian}
We consider the ground-state spin manifold of a single, negatively-charged germanium-vacancy (GeV) center, consisting of an electron spin $S=1/2$ coupled to a nuclear spin $I=1/2$. 
We work in units where $\hbar = 1$ and define spin operators as $S_\alpha = \tfrac{1}{2} \sigma_\alpha \otimes \mathbb{1}$ and $I_\alpha  = \tfrac{1}{2} \mathbb{1} \otimes \sigma_\alpha$ with Pauli matrices $\sigma_\alpha$.
A static magnetic field is applied along the principal axis of the GeV center, which we define as the laboratory $z$-axis, determining the quantization direction.
In this frame, the noise-free driven Hamiltonian reads
\begin{align}
H = -\omega_e S_z - \omega_I I_z + A_{zz} S_z I_z + A_{zx} S_z I_x + \Omega(t) S_x ,
\end{align}
where $\omega_e$ and $\omega_I$ are the electron and nuclear Larmor frequencies, respectively.
The effective hyperfine interaction is truncated to the dominant secular and symmetry-allowed terms $A_{zz}$ and $A_{zx}$ \cite{grimm2025coherent}.
To simplify the driven dynamics, we move to a rotating frame with respect to the electron Zeeman term using
\begin{equation}
R(t) = \exp(-i \omega_e t S_z),
\end{equation}
and assume a microwave control field of the form
\begin{equation}
\Omega(t) = 2 \Omega_{\text{MW}}(t) \cos(\omega_d t - \phi),
\end{equation}
where the drive frequency $\omega_d = \omega_e + \Delta$ is detuned by $\Delta$ from resonance and $\phi$ denotes the control phase.
In the investigated two-qubit system,  we used for implementing the CNOT gates the $\ket{\downarrow \uparrow} \longleftrightarrow \ket{\uparrow \uparrow}$ transition frequency as drive frequency, while for the SWAP gate we used the frequency of the $\ket{\uparrow \downarrow} \longleftrightarrow \ket{\downarrow \uparrow}$ transition.
Applying the transformation $H' = R(t) H R^\dagger(t) - i R(t)\dot{R}^\dagger(t)$ and retaining only slow terms within the rotating-wave approximation (RWA), we obtain the effective control Hamiltonian
\begin{align}
H_c' &= R(t) H_c(t) R^\dagger(t) \\
&= \Omega_{\text{MW}}(t)
\bigl[\cos(\Delta t - \phi) S_x - \sin(\Delta t - \phi) S_y\bigr].
\end{align}
In the optimal control simulations, both the microwave envelope $\Omega_{\text{MW}}(t)$ and the phase $\phi(t)$ are treated as time-dependent control parameters.
\section{Open-system dynamics and decoherence}
\label{sec:open-system}
Dephasing noise is incorporated via a stochastic detuning term added to the electron Zeeman splitting,
\begin{equation}
\omega_e \rightarrow \omega_e + \beta(t),
\end{equation}
corresponding to an additional Hamiltonian contribution $\beta(t) S_z$.
The noise process $\beta(t)$ is modeled as an Ornstein--Uhlenbeck (OU) process, capturing non-Markovian dephasing with finite correlation time.
The OU process is characterized by its noise strength $\sigma$ and correlation time $t_c$, which we choose such that the resulting dynamics reproduce the experimentally measured decoherence times $T_2^*$ and $T_2$:
\begin{equation}
\sigma = \frac{\sqrt{2}}{2 T_2^*},
\qquad
t_c = \frac{T_2^3}{6 (T_2^*)^2},
\end{equation}
where $T_2^* = \SI{1.542}{\micro\second}$ and $T_2 = \SI{605}{\micro\second}$.\\
The noise is discretized using the exact update rule
\begin{align}
\beta(t + \Delta t)
= e^{-\Delta t / t_c} \beta(t)
+ \sigma \sqrt{1 - e^{-2\Delta t / t_c}}\, \xi_n,
\end{align}
where $\xi_n \sim \mathcal{N}(0,1)$ are independent Gaussian random variables.
Initial values are drawn from the stationary distribution
\begin{equation}
\beta(0) \sim \mathcal{N}(0,\sigma^2).
\end{equation}
The time evolution is carried out via first-order Trotterization with time step $\Delta t = \SI{1}{\nano\second}$.
All reported fidelities are obtained by averaging over an ensemble of noise realizations.
\section{Numerical methods and optimal control}
\label{sec:num_methods}
The optimal control pulses were obtained using the open-source
optimal control suite QuOCS ~\cite{rossignolo2023quocs}, a Python framework designed for pulse optimization in quantum systems.
The optimization employed the dCRAB algorithm with Nelder-Mead as search algorithm.
It used 10 superiterations each consisting of $N_{\text{eval}}$ = 5000 FoM evaluations.
The control fields were expanded in a truncated
Fourier basis consisting of 20 basis functions.
The corresponding
frequencies were sampled uniformly from the interval
$\omega t_f \in [0.1,40]$, meaning the highest possible frequency allowed the pulse to swing 40 times in the total evolution time $t_f$.
This approach is compatible with existing control hardware. 
To ensure smooth switch-on and switch-off of the control fields, all
pulses were multiplied by a time-dependent Gaussian envelope function $s(t)$,
defined as
\begin{equation}
s(t)=
\begin{cases}
\alpha\!\left[
e^{-\frac{(t-t_r)^2}{2\sigma^2}}
-
e^{-\frac{t_r^2}{2\sigma^2}}
\right], & 0<t\le t_r, \\[0.8em]

1, & t_r<t<t_f-t_r, \\[0.8em]

\alpha\!\left[
e^{-\frac{(t-(t_f-t_r))^2}{2\sigma^2}}
-
e^{-\frac{t_r^2}{2\sigma^2}}
\right], & t_f-t_r\le t<t_f .
\end{cases}
\end{equation}
Here, $\alpha = \left[1-\exp\!\left(-\frac{t_r^2}{2\sigma^2}\right)\right]^{-1}$
is a normalization constant, $t_r=\SI{100}{\nano \s}$ denotes the rise time, and $\sigma = t_r/4$ sets the width of the Gaussian ramps.
\section{Hadamard Gates}
\label{sec:Hadarmards}
\begin{figure}[htb]
    \centering
    \includegraphics{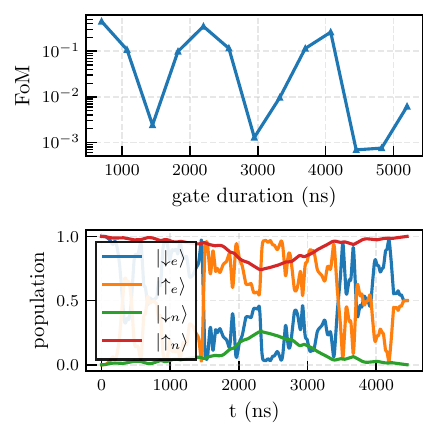}
    \caption{Optimization process of a Hadamard ($V_G = \mathcal{H} \otimes \mathbb{1}$) gate on the electron spin, the upper panel shows the reached FoM for different gate durations, the best reached fidelity was  $\Fav(V_G) =\SI{99.927 \pm 0.005}\%$ at gate duration of $\SI{4.4450}{\micro \s}$ and averaged over $N=5000$ noise realizations. The lower panel shows the populations resulting from the optimized pulse with gate duration of ($\SI{4.450}{\micro \s}$) while the other spin is traced out.}
    \label{fig:H_E}
\end{figure}
\begin{figure}
    \centering
    \includegraphics{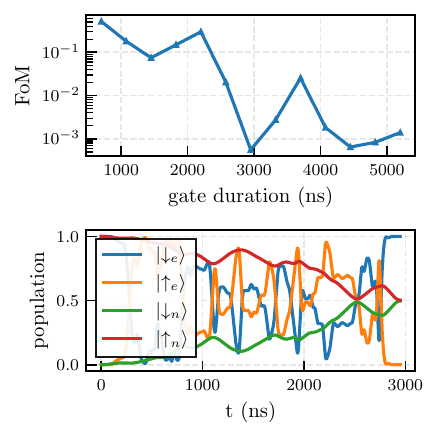}
    \caption{Optimization process of a Hadamard ($V_G = \mathbb{1} \otimes \mathcal{H}$) gate on the nuclear spin, the upper panel shows the reached FoM for different gate durations, the best reached fidelity was  $\Fav(V_G) =\SI{99.942 \pm 0.003}\%$ at gate duration of $\SI{2.950}{\micro \s}$ and averaged over $N=5000$ noise realizations. The lower panel shows the populations resulting from the optimized pulse with gate duration of ($\SI{5.2}{\micro \s}$) while the other spin is traced out.}
    \label{fig:H_N}
\end{figure}
While in this article we focus on two-qubit gates, we also need Hadamard gates for many protocols.
On the one hand, these pose challenges in the implementation due to the interaction terms. On the other hand, the interaction allows us to generate a fast Hadamard gate also on the nuclear spin through driving of the electron spin. 
In \autoref{fig:H_E} we show the results for an optimized Hadamard gate on the electron spin and in \autoref{fig:H_N} we show them for the optimized Hadamard gate on the nuclear spin. 
We can see from the figures that in both cases for a gate duration similar to the gate duration of the two-qubit gates, we can find gate fidelties beyond $99.9\,\% $, also comparable to the two-qubit gate fidelities.
\section{Remarks on the Cartan decomposition}
\label{sec:remarks}
Any two-qubit unitary $U \in \mathrm{SU}(4)$ can be decomposed into local and nonlocal parts according to the Cartan (or KAK) decomposition,
\begin{equation}
U = K_1 A K_2,
\end{equation}
where $K_1$ and $K_2$ are local unitaries and the nonlocal component takes the canonical form
\begin{equation}
A = \exp\!\left[-\frac{i}{2}
\left(c_1\, XX + c_2\, YY + c_3\, ZZ\right)\right].
\end{equation}
In particular, gates that are locally equivalent to the SWAP operation satisfy
\begin{equation}
U = e^{i\phi}\, k_1 \, \mathrm{SWAP}\, k_2,
\end{equation}
where $k_1$ and $k_2$ are local unitaries and $\phi$ is a global phase.
Multiplying from the right by $\mathrm{SWAP}$ yields
\begin{align}
U\,\mathrm{SWAP}
&= e^{i\phi}\, k_1 \, \mathrm{SWAP}\, k_2 \, \mathrm{SWAP} 
\end{align}
Since conjugation by $\mathrm{SWAP}$ maps local unitaries to local unitaries, the operator
\begin{equation}
k \equiv e^{i\phi}\, k_1 \, (\mathrm{SWAP}\, k_2 \, \mathrm{SWAP})
\end{equation}
is itself local.
Hence, any SWAP-equivalent unitary can be written as
\begin{equation}
k = U\,\mathrm{SWAP}.
\end{equation}
Assuming factorizations $k_1 = u_1 \otimes v_1$ and $k_2 = u_2 \otimes v_2$, and using
\begin{equation}
\mathrm{SWAP}(A \otimes B)\mathrm{SWAP} = B \otimes A,
\end{equation}
we obtain, up to a global phase,
\begin{equation}
k = (u_1 v_2) \otimes (v_1 u_2),
\end{equation}
demonstrating explicitly that $U\,\mathrm{SWAP}$ is local.
\subsection{Relation between U$_{\text{ideal}}$ and SWAP}
For verifying that the numerically optained nonlocal unitary $U_{\text{ideal}}$ is locally equivalent to a SWAP gate, we analyze the operator
\begin{align}
    K = U_{\text{ideal}}^\dagger \text{SWAP}.
\end{align}
If $U_{\text{ideal}}$ differs from SWAP only by local unitaries, i.e. $\mathcal{F}_{\text{nl}}(\text{SWAP}, U_{\text{ideal}}) = 1$, K must factorize as 
\begin{equation}
K = K_1 \otimes K_2,
\quad
K_j \in \mathrm{SU}(2).
\end{equation}
We compute the closest product unitary $K_1 \otimes K_2$ in Frobenius norm using an operator Schmidt rank-1 approximation. That is, we minimize the cost function
\begin{align}
\label{eq:cost}
    \mathcal{C} = \frac{1}{d^2}\abs{\Tr{K^\dagger (K_1\otimes K_2)}}^2 
\end{align}
and write the resulting single–qubit unitaries  as
\begin{equation}
\label{eq:singleK}
K_j = \exp(-i\, \vec n_j \cdot \vec \sigma).
\end{equation}
The numerical results are summarized in \autoref{tab:swapvalues}.
\begin{table}[ht]
\centering
\caption{SWAP alternatives. Here, T denotes the gate duration. $1-F_\text{nl}$ is the non-local gate infidelity from Eq.~\eqref{eq:fnl}, $\mathcal{C}$ is the cost function introduced in Eq.~\eqref{eq:cost}, and $\vec{n}_{1}, \vec{n}_2$ are the rotation vectors for the single-qubit rotations in Eq.~\eqref{eq:singleK}.}
\begin{tabular}{|c| c| c| c| c|}
\hline
$T$ &
$1-\mathcal F_{\mathrm{nl}}$ &
$\mathcal C$ &
$\vec n_1$ &
$\vec n_2$ \\
\hline
$\SI{2.950}{\micro\second}$ &
$2.3\times 10^{-6}$ &
$4.56\times 10^{-6}$ &
$\begin{pmatrix}
0.5142\\
0.9110\\
-0.8265
\end{pmatrix}$ &
$\begin{pmatrix}
0.7688\\
-0.4076\\
-0.4420
\end{pmatrix}$ \\
\hline
$\SI{4.825}{\micro\second}$ &
$3.25\times 10^{-8}$ &
$1.51\times 10^{-6}$ &
$\begin{pmatrix}
0.4817\\
1.6932\\
-0.7045
\end{pmatrix}$ &
$\begin{pmatrix}
-0.9399\\
0.4227\\
1.4484
\end{pmatrix}$ \\
\hline
\end{tabular}
\label{tab:swapvalues}
\end{table}
\bibliographystyle{apsrev4-2}
\bibliography{references} 

@article{Takou2023,
  title = {Precise Control of Entanglement in Multinuclear Spin Registers Coupled to Defects},
  author = {Takou, Evangelia and Barnes, Edwin and Economou, Sophia E.},
  journal = {Phys. Rev. X},
  volume = {13},
  issue = {1},
  pages = {011004},
  numpages = {40},
  year = {2023},
  month = {Jan},
  publisher = {American Physical Society},
  doi = {10.1103/PhysRevX.13.011004},
  url = {https://link.aps.org/doi/10.1103/PhysRevX.13.011004}
}

@article{Zeier2007,
  title = {Time-optimal synthesis of unitary transformations in a coupled fast and slow qubit system},
  author = {Zeier, Robert and Yuan, Haidong and Khaneja, Navin},
  journal = {Phys. Rev. A},
  volume = {77},
  issue = {3},
  pages = {032332},
  numpages = {8},
  year = {2008},
  month = {Mar},
  publisher = {American Physical Society},
  doi = {10.1103/PhysRevA.77.032332},
  url = {https://link.aps.org/doi/10.1103/PhysRevA.77.032332}
}

@misc{wang2025,
      title={Circuit Design for a Star-shaped Spin-Qubit Processor via Algebraic Decomposition and Optimal Control}, 
      author={Yaqing X. Wang and Tommaso Calarco and Felix Motzoi and Matthias M. Müller},
      year={2025},
      eprint={2506.16900},
      archivePrefix={arXiv},
      primaryClass={quant-ph},
      url={https://arxiv.org/abs/2506.16900}, 
}

@article{Chen2020,
	doi = {10.1088/1367-2630/abb0fb},
	url = {https://dx.doi.org/10.1088/1367-2630/abb0fb},
	year = {2020},
	month = {sep},
	publisher = {IOP Publishing},
	volume = {22},
	number = {9},
	pages = {093068},
	author = {YunHeng Chen and Sophie Stearn and Scott Vella and Andrew Horsley and Marcus W Doherty},
	title = {Optimisation of diamond quantum processors},
	journal = {New Journal of Physics}
}

@article{Khaneja2001cartan,
  title={Cartan decomposition of SU (2n) and control of spin systems},
  author={Khaneja, Navin and Glaser, Steffen J},
  journal={Chemical Physics},
  volume={267},
  number={1-3},
  pages={11--23},
  year={2001},
  publisher={Elsevier}
}

@article{Zhang2004a,
	title = {Optimal quantum circuit synthesis from controlled-unitary gates},
	author = {Zhang, Jun and Vala, Jiri and Sastry, Shankar and Whaley, K. Birgitta},
	journal = {Phys. Rev. A},
	volume = {69},
	issue = {4},
	pages = {042309},
	numpages = {6},
	year = {2004},
	month = {Apr},
	publisher = {American Physical Society},
	doi = {10.1103/PhysRevA.69.042309},
}

@article{Zhang2003b,
	title = {Exact Two-Qubit Universal Quantum Circuit},
	author = {Zhang, Jun and Vala, Jiri and Sastry, Shankar and Whaley, K. Birgitta},
	journal = {Phys. Rev. Lett.},
	volume = {91},
	issue = {2},
	pages = {027903},
	numpages = {4},
	year = {2003},
	month = {Jul},
	publisher = {American Physical Society},
	doi = {10.1103/PhysRevLett.91.027903},
}

@article{Nakajima2005ANA,
  title={A new algorithm for producing quantum circuits using KAK decompositions},
  author={Yumi Nakajima and Yasuhito Kawano and Hiroshi Sekigawa},
  journal={Quantum Inf. Comput.},
  year={2005},
  volume={6},
  pages={67-80},
  url={https://api.semanticscholar.org/CorpusID:1337965}
}

@article{Yuan2015,
   author = {Haidong Yuan and Robert Zeier and Nikolas Pomplun and Steffen J. Glaser and Navin Khaneja},
   doi = {10.1103/PhysRevA.92.053414},
   issn = {10941622},
   issue = {5},
   journal = {Physical Review A - Atomic, Molecular, and Optical Physics},
   month = {11},
   publisher = {American Physical Society},
   title = {Time-optimal polarization transfer from an electron spin to a nuclear spin},
   volume = {92},
   year = {2015},
}

@article{Frank2017,
	Author = {Frank, Florian and Unden, Thomas and Zoller, Jonathan and Said, Ressa S. and Calarco, Tommaso and Montangero, Simone and Naydenov, Boris and Jelezko, Fedor},
	Da = {2017/12/01},
	Doi = {10.1038/s41534-017-0049-8},
	Id = {Frank2017},
	Isbn = {2056-6387},
	Journal = {npj Quantum Information},
	Number = {1},
	Pages = {48},
	Title = {Autonomous calibration of single spin qubit operations},
	Ty = {JOUR},
	Volume = {3},
	Year = {2017}}

@article{Said2009,
  title = {Robust control of entanglement in a nitrogen-vacancy center coupled to a $^{13}\text{C}$ nuclear spin in diamond},
  author = {Said, R. S. and Twamley, J.},
  journal = {Phys. Rev. A},
  volume = {80},
  issue = {3},
  pages = {032303},
  numpages = {7},
  year = {2009},
  month = {Sep},
  publisher = {American Physical Society},
  doi = {10.1103/PhysRevA.80.032303},
  url = {https://link.aps.org/doi/10.1103/PhysRevA.80.032303}
}

@article{Oshnik2022,
		title = {Robust magnetometry with single nitrogen-vacancy centers via two-step optimization},
		author = {Oshnik, Nimba and Rembold, Phila and Calarco, Tommaso and Montangero, Simone and Neu, Elke and M\"uller, Matthias M.},
		journal = {Phys. Rev. A},
		volume = {106},
		issue = {1},
		pages = {013107},
		numpages = {15},
		year = {2022},
		month = {Jul},
		publisher = {American Physical Society},
		doi = {10.1103/PhysRevA.106.013107},
		url = {https://link.aps.org/doi/10.1103/PhysRevA.106.013107}
	}

@article{lim2025,
  title = {Efficiency of optimal control for noisy spin qubits in diamond},
  author = {Lim, Hendry M. and Genov, Genko T. and Sailer, Roberto and Fahrurrachman, Alfaiz and Majidi, Muhammad A. and Jelezko, Fedor and Said, Ressa S.},
  journal = {Phys. Rev. Appl.},
  volume = {24},
  issue = {5},
  pages = {054064},
  numpages = {20},
  year = {2025},
  month = {Nov},
  publisher = {American Physical Society},
  doi = {10.1103/3hnz-bysr},
  url = {https://link.aps.org/doi/10.1103/3hnz-bysr}
}

@article{knaut2024entanglement,
  title={Entanglement of nanophotonic quantum memory nodes in a telecom network},
  author={Knaut, Can M and Suleymanzade, Aziza and Wei, Y-C and Assumpcao, Daniel R and Stas, P-J and Huan, Yan Qi and Machielse, Bartholomeus and Knall, Erik N and Sutula, Madison and Baranes, Gefen and others},
  journal={Nature},
  volume={629},
  number={8012},
  pages={573--578},
  year={2024},
  publisher={Nature Publishing Group UK London}
}

@misc{frey2026,
	title={Protocols for Entangling Gates for Group IV Color-Centers in Diamond}, 
	author={Frey,Jurek and Wilhelm, Frank K. and Müller, Matthias M.},
	year={2026},
	jounrall={in preparation} 
}

@misc{baran2026,
      author={Bora Baran and Tommaso Calarco and Matthias M. Müller and Felix Motzoi},
      year={2026},
      eprint={2602.18375},
      archivePrefix={arXiv},
      primaryClass={quant-ph},
      url={https://arxiv.org/abs/2602.18375}, 
}

@article{Klotz2026,
	title = {Bipartite entanglement in a nuclear spin register mediated by a quasi-free electron spin},
	volume = {17},
	issn = {2041-1723},
	url = {https://doi.org/10.1038/s41467-026-70154-3},
	doi = {10.1038/s41467-026-70154-3},
	number = {1},
	journal = {Nature Communications},
	author = {Klotz, Marco and Tangemann, Andreas and Opferkuch, David and Kubanek, Alexander},
	month = mar,
	year = {2026},
	pages = {2325},
}

@misc{koch2025,
	title={Introduction to quantum control: From basic concepts to applications in quantum technologies}, 
	author={Christiane P. Koch},
	year={2025},
	eprint={2512.04990},
	archivePrefix={arXiv},
	primaryClass={quant-ph},
	url={https://arxiv.org/abs/2512.04990} 
}

@article{rembold2020,
author = {Rembold,Phila  and Oshnik,Nimba  and Müller,Matthias M.  and Montangero,Simone  and Calarco,Tommaso  and Neu,Elke },
title = {Introduction to quantum optimal control for quantum sensing with nitrogen-vacancy centers in diamond},
journal = {AVS Quantum Science},
volume = {2},
number = {2},
pages = {024701},
year = {2020},
doi = {10.1116/5.0006785},
URL = {https://doi.org/10.1116/5.0006785},
Eprint = { https://doi.org/10.1116/5.0006785}
}

@article{muller2022one,
  title={One decade of quantum optimal control in the chopped random basis},
  author={M{\"u}ller, Matthias M and Said, Ressa S and Jelezko, Fedor and Calarco, Tommaso and Montangero, Simone},
  journal={Reports on progress in physics},
  volume={85},
  number={7},
  pages={076001},
  year={2022},
  publisher={IOP Publishing}
}

@article{vetter2024gate,
  title={Gate-set evaluation metrics for closed-loop optimal control on nitrogen-vacancy center ensembles in diamond},
  author={Vetter, Philipp J and Reisser, Thomas and Hirsch, Maximilian G and Calarco, Tommaso and Motzoi, Felix and Jelezko, Fedor and M{\"u}ller, Matthias M},
  journal={npj Quantum Information},
  volume={10},
  number={1},
  pages={96},
  year={2024},
  publisher={Nature Publishing Group UK London}
}

@article{rossignolo2023quocs,
    title = {QuOCS: The quantum optimal control suite},
    journal = {Computer Physics Communications},
    volume = {291},
    pages = {108782},
    year = {2023},
    issn = {0010-4655},
    doi = {https://doi.org/10.1016/j.cpc.2023.108782},
    url = {https://www.sciencedirect.com/science/article/pii/S0010465523001273},
    author = {Marco Rossignolo and others}
}

@article{uhlenbeck1930theory,
  title={On the theory of the Brownian motion},
  author={Uhlenbeck, George E and Ornstein, Leonard S},
  journal={Physical review},
  volume={36},
  number={5},
  pages={823},
  year={1930},
  publisher={APS}
}

@article{bradac2019quantum,
  title={Quantum nanophotonics with group IV defects in diamond},
  author={Bradac, Carlo and Gao, Weibo and Forneris, Jacopo and Trusheim, Matthew E and Aharonovich, Igor},
  journal={Nature communications},
  volume={10},
  number={1},
  pages={5625},
  year={2019},
  publisher={Nature Publishing Group UK London}
}

@article{ruf2021quantum,
  title={Quantum networks based on color centers in diamond},
  author={Ruf, Maximilian and Wan, Noel H and Choi, Hyeongrak and Englund, Dirk and Hanson, Ronald},
  journal={Journal of Applied Physics},
  volume={130},
  number={7},
  year={2021},
  publisher={AIP Publishing}
}

@article{stas2025entanglement,
  title={Entanglement Assisted Non-local Optical Interferometry in a Quantum Network},
  author={Stas, P-J and Wei, Y-C and Sirotin, M and Huan, YQ and Yazlar, U and Arias, F Abdo and Knyazev, E and Baranes, G and Machielse, B and Grandi, S and others},
  journal={arXiv preprint arXiv:2509.09464},
  year={2025}
}

@article{vetter2025room,
  title={Room-Temperature Quantum Simulation with Atomically Thin Nuclear Spin Layers in Diamond},
  author={Vetter, Philipp J and Findler, Christoph and Verd{\'u}, Antonio and Kost, Matthias and Blinder, R{\'e}mi and Fuhrmann, Jens and Osterkamp, Christian and Lang, Johannes and Plenio, Martin B and Prior, Javier and others},
  journal={arXiv preprint arXiv:2510.27374},
  year={2025}
}

@article{wei2025universal,
  title={Universal distributed blind quantum computing with solid-state qubits},
  author={Wei, Y-C and Stas, P-J and Suleymanzade, Aziza and Baranes, Gefen and Machado, Francisco and Huan, Yan Qi and Knaut, Can M and Ding, Sophie W and Merz, Moritz and Knall, Erik N and others},
  journal={Science},
  volume={388},
  number={6746},
  pages={509--513},
  year={2025},
  publisher={American Association for the Advancement of Science}
}

@article{senkalla2024germanium,
  title = {Germanium Vacancy in Diamond Quantum Memory Exceeding 20 ms},
  author = {Senkalla, Katharina and Genov, Genko and Metsch, Mathias H. and Siyushev, Petr and Jelezko, Fedor},
  journal = {Phys. Rev. Lett.},
  volume = {132},
  issue = {2},
  pages = {026901},
  numpages = {6},
  year = {2024},
  month = {Jan},
  publisher = {American Physical Society},
  doi = {10.1103/PhysRevLett.132.026901},
  url = {https://link.aps.org/doi/10.1103/PhysRevLett.132.026901}
}

@misc{Gundlapalli2025,
      title={High-Fidelity Single-Shot Readout and Selective Nuclear Spin Control for a Spin-1/2 Quantum Register in Diamond}, 
      author={Prithvi Gundlapalli and Philipp J. Vetter and Genko Genov and Michael Olney-Fraser and Peng Wang and Matthias M. Müller and Katharina Senkalla and Fedor Jelezko},
      year={2025},
      eprint={2510.09164},
      archivePrefix={arXiv},
      primaryClass={quant-ph},
      url={https://arxiv.org/abs/2510.09164}, 
}

@article{grimm2025coherent,
  title={Coherent control of a long-lived nuclear memory spin in a germanium-vacancy multi-qubit node},
  author={Grimm, Nick and Senkalla, Katharina and Vetter, Philipp J and Frey, Jurek and Gundlapalli, Prithvi and Calarco, Tommaso and Genov, Genko and M{\"u}ller, Matthias M and Jelezko, Fedor},
  journal={Physical Review Letters},
  volume={134},
  number={4},
  pages={043603},
  year={2025},
  publisher={APS}
}

@article{rach2015dressing,
  title={Dressing the chopped-random-basis optimization: A bandwidth-limited access to the trap-free landscape},
  author={Rach, Norman and M{\"u}ller, Matthias M and Calarco, Tommaso and Montangero, Simone},
  journal={Physical Review A},
  volume={92},
  number={6},
  pages={062343},
  year={2015},
  publisher={APS}
}

@article{egger2014adaptive,
  title={Adaptive hybrid optimal quantum control for imprecisely characterized systems},
  author={Egger, Daniel J and Wilhelm, Frank K},
  journal={Physical Review Letters},
  volume={112},
  number={24},
  pages={240503},
  year={2014},
  publisher={APS}
}

@article{taminiau2014universal,
  title={Universal control and error correction in multi-qubit spin registers in diamond},
  author={Taminiau, Tim H and Cramer, J and van der Sar, T and Dobrovitski, VV and Hanson, R},
  journal={Nature Nanotechnology},
  volume={9},
  number={3},
  pages={171--176},
  year={2014},
  publisher={Nature Publishing Group}
}

@article{bradley2019ten,
  title={A ten-qubit solid-state spin register with quantum memory up to one minute},
  author={Bradley, Conor E and Randall, Joe and Abobeih, Mohamed H and Berrevoets, Remon C and Degen, Maarten J and Bakker, Michiel A and Markham, Matthew and Twitchen, Daniel J and Taminiau, Tim H},
  journal={Physical Review X},
  volume={9},
  number={3},
  pages={031045},
  year={2019},
  publisher={APS}
}

@article{resch2025high,
  title = {High-Fidelity Control of a {$^{13}\text{C}$} Nuclear Spin Coupled to a Tin-Vacancy Center in Diamond},
  author = {Resch, Jeremias and Karapatzakis, Ioannis and Elshorbagy, Mohamed and Schrodin, Marcel and Fuchs, Philipp and Gra\ss{}hoff, Philipp and Kussi, Luis and S\"urgers, Christoph and Popov, Cyril and Becher, Christoph and Wernsdorfer, Wolfgang and Hunger, David},
  journal = {Phys. Rev. X},
  volume = {16},
  issue = {1},
  pages = {011060},
  numpages = {18},
  year = {2026},
  month = {Mar},
  publisher = {American Physical Society},
  doi = {10.1103/bmc6-qvwq},
  url = {https://link.aps.org/doi/10.1103/bmc6-qvwq}
}

@article{pompili21realization,
author = {M. Pompili  and S. L. N. Hermans  and S. Baier  and H. K. C. Beukers  and P. C. Humphreys  and R. N. Schouten  and R. F. L. Vermeulen  and M. J. Tiggelman  and L. dos Santos Martins  and B. Dirkse  and S. Wehner  and R. Hanson },
title = {Realization of a multinode quantum network of remote solid-state qubits},
journal = {Science},
volume = {372},
number = {6539},
pages = {259-264},
year = {2021},
doi = {10.1126/science.abg1919},
URL = {https://www.science.org/doi/abs/10.1126/science.abg1919},
eprint = {https://www.science.org/doi/pdf/10.1126/science.abg1919}}

@article{kalb17entanglement,
author = {N. Kalb  and A. A. Reiserer  and P. C. Humphreys  and J. J. W. Bakermans  and S. J. Kamerling  and N. H. Nickerson  and S. C. Benjamin  and D. J. Twitchen  and M. Markham  and R. Hanson },
title = {Entanglement distillation between solid-state quantum network nodes},
journal = {Science},
volume = {356},
number = {6341},
pages = {928-932},
year = {2017},
doi = {10.1126/science.aan0070},
URL = {https://www.science.org/doi/abs/10.1126/science.aan0070},
eprint = {https://www.science.org/doi/pdf/10.1126/science.aan0070}}

@article{stas22robust,
author = {P.-J. Stas  and Y. Q. Huan  and B. Machielse  and E. N. Knall  and A. Suleymanzade  and B. Pingault  and M. Sutula  and S. W. Ding  and C. M. Knaut  and D. R. Assumpcao  and Y.-C. Wei  and M. K. Bhaskar  and R. Riedinger  and D. D. Sukachev  and H. Park  and M. Lončar  and D. S. Levonian  and M. D. Lukin },
title = {Robust multi-qubit quantum network node with integrated error detection},
journal = {Science},
volume = {378},
number = {6619},
pages = {557-560},
year = {2022},
doi = {10.1126/science.add9771},
URL = {https://www.science.org/doi/abs/10.1126/science.add9771},
eprint = {https://www.science.org/doi/pdf/10.1126/science.add9771}}

@article{watts2015optimizing,
  title={Optimizing for an arbitrary perfect entangler. I. Functionals},
  author={Watts, Paul and Vala, Ji{\v{r}}{\'\i} and M{\"u}ller, Matthias M and Calarco, Tommaso and Whaley, K Birgitta and Reich, Daniel M and Goerz, Michael H and Koch, Christiane P},
  journal={Physical Review A},
  volume={91},
  number={6},
  pages={062306},
  year={2015},
  publisher={APS}
}

@article{Mueller2011,
	Author = {M\"uller, M. M. and Reich, D. M. and Murphy, M. and Yuan, H. and Vala, J. and Whaley, K. B. and Calarco, T. and Koch, C. P.},
	Doi = {10.1103/PhysRevA.84.042315},
	Issue = {4},
	Journal = {Phys. Rev. A},
	Month = {Oct},
	Numpages = {8},
	Pages = {042315},
	Publisher = {American Physical Society},
	Title = {Optimizing entangling quantum gates for physical systems},
	Volume = {84},
	Year = {2011}}

@article{Goerz2015,
	Author = {Goerz, Michael H. and Gualdi, Giulia and Reich, Daniel M. and Koch, Christiane P. and Motzoi, Felix and Whaley, K. Birgitta and Vala, Jiri and M\"uller, Matthias M. and Montangero, Simone and Calarco, Tommaso},
	Doi = {10.1103/PhysRevA.91.062307},
	Issue = {6},
	Journal = {Phys. Rev. A},
	Month = {Jun},
	Numpages = {11},
	Pages = {062307},
	Publisher = {American Physical Society},
	Title = {Optimizing for an arbitrary perfect entangler. II. Application},
	Volume = {91},
	Year = {2015}}

@article{Goerz2017,
	title = {Charting the circuit {QED} design landscape using optimal control theory},
	volume = {3},
	issn = {2056-6387},
	url = {https://doi.org/10.1038/s41534-017-0036-0},
	doi = {10.1038/s41534-017-0036-0},
	number = {1},
	journal = {npj Quantum Information},
	author = {Goerz, Michael H. and Motzoi, Felix and Whaley, K. Birgitta and Koch, Christiane P.},
	month = sep,
	year = {2017},
	pages = {37},
}

@article{childress2006coherent,
  title={Coherent dynamics of coupled electron and nuclear spin qubits in diamond},
  author={Childress, L and Gurudev Dutt, MV and Taylor, JM and Zibrov, AS and Jelezko, F and Wrachtrup, J and Hemmer, PR and Lukin, MD},
  journal={Science},
  volume={314},
  number={5797},
  pages={281--285},
  year={2006},
  publisher={American Association for the Advancement of Science}
}

@article{beukers2025control,
  title={Control of solid-state nuclear spin qubits using an electron spin-1/2},
  author={Beukers, Hans KC and Waas, Christopher and Pasini, Matteo and Van Ommen, Hendrik B and Ademi, Zarije and Iuliano, Mariagrazia and Codreanu, Nina and Brevoord, Julia M and Turan, Tim and Taminiau, Tim H and others},
  journal={Physical Review X},
  volume={15},
  number={2},
  pages={021011},
  year={2025},
  publisher={APS}
}

@article{glaser2015training,
  title={Training Schr{\"o}dinger’s cat: Quantum optimal control: Strategic report on current status, visions and goals for research in Europe},
  author={Glaser, Steffen J and Boscain, Ugo and Calarco, Tommaso and Koch, Christiane P and K{\"o}ckenberger, Walter and Kosloff, Ronnie and Kuprov, Ilya and Luy, Burkhard and Schirmer, Sophie and Schulte-Herbr{\"u}ggen, Thomas and others},
  journal={The European Physical Journal D},
  volume={69},
  number={12},
  pages={279},
  year={2015},
  publisher={Springer}
}

@article{dolde2014high,
  title={High-fidelity spin entanglement using optimal control},
  author={Dolde, Florian and Bergholm, Ville and Wang, Ya and Jakobi, Ingmar and Naydenov, Boris and Pezzagna, S{\'e}bastien and Meijer, Jan and Jelezko, Fedor and Neumann, Philipp and Schulte-Herbr{\"u}ggen, Thomas and others},
  journal={Nature communications},
  volume={5},
  number={1},
  pages={3371},
  year={2014},
  publisher={Nature Publishing Group UK London}
}

@article{dobrovitski2013quantum,
  title={Quantum control over single spins in diamond},
  author={Dobrovitski, VV and Fuchs, GD and Falk, AL and Santori, C and Awschalom, DD},
  journal={Annu. Rev. Condens. Matter Phys.},
  volume={4},
  number={1},
  pages={23--50},
  year={2013},
  publisher={Annual Reviews}
}

@article{khodjasteh2009dynamical,
  title = {Dynamical quantum error correction of unitary operations with bounded controls},
  author = {Khodjasteh, K. and Viola, L.},
  journal = {Physical Review A},
  volume = {80},
  pages = {032314},
  year = {2009},
  doi = {10.1103/PhysRevA.80.032314}
}

@article{robledo2011high,
  title={High-fidelity projective read-out of a solid-state spin quantum register},
  author={Robledo, Lucio and Childress, Lilian and Bernien, Hannes and Hensen, Bas and Alkemade, Paul FA and Hanson, Ronald},
  journal={Nature},
  volume={477},
  number={7366},
  pages={574--578},
  year={2011},
  publisher={Nature Publishing Group UK London}
}

@article{hegde2020efficient,
  title={Efficient quantum gates for individual nuclear spin qubits by indirect control},
  author={Hegde, Swathi S and Zhang, Jingfu and Suter, Dieter},
  journal={Physical review letters},
  volume={124},
  number={22},
  pages={220501},
  year={2020},
  publisher={APS}
}

@article{vallabhapurapu2022indirect,
  title={Indirect control of the 29 SiV- nuclear spin in diamond},
  author={Vallabhapurapu, Hyma H and Adambukulam, Chris and Saraiva, Andre and Laucht, Arne},
  journal={Physical Review B},
  volume={105},
  number={20},
  pages={205435},
  year={2022},
  publisher={APS}
}

@article{zhang2003geometric,
  title={Geometric theory of nonlocal two-qubit operations},
  author={Zhang, Jun and Vala, Jiri and Sastry, Shankar and Whaley, K Birgitta},
  journal={Physical Review A},
  volume={67},
  number={4},
  pages={042313},
  year={2003},
  publisher={APS}
}

@article{makhlin2002nonlocal,
  title={Nonlocal properties of two-qubit gates and mixed states, and the optimization of quantum computations},
  author={Makhlin, Yuriy},
  journal={Quantum Information Processing},
  volume={1},
  number={4},
  pages={243--252},
  year={2002},
  publisher={Springer}
}

@article{kraus2001optimal,
  title={Optimal creation of entanglement using a two-qubit gate},
  author={Kraus, Barbara and Cirac, J Ignacio},
  journal={Physical Review A},
  volume={63},
  number={6},
  pages={062309},
  year={2001},
  publisher={APS}
}

@article{van2012decoherence,
  title={Decoherence-protected quantum gates for a hybrid solid-state spin register},
  author={Van der Sar, T and Wang, ZH and Blok, MS and Bernien, H and Taminiau, TH and Toyli, DM and Lidar, DA and Awschalom, DD and Hanson, Robin and Dobrovitski, VV},
  journal={Nature},
  volume={484},
  number={7392},
  pages={82--86},
  year={2012},
  publisher={Nature Publishing Group UK London}
}

@article{scheuer2014precise,
  title={Precise qubit control beyond the rotating wave approximation},
  author={Scheuer, Jochen and Kong, Xi and Said, Ressa S and Chen, Jeson and Kurz, Andrea and Marseglia, Luca and Du, Jiangfeng and Hemmer, Philip R and Montangero, Simone and Calarco, Tommaso and others},
  journal={New Journal of Physics},
  volume={16},
  number={9},
  pages={093022},
  year={2014},
  publisher={IOP Publishing}
}

@article{jiang2009repetitive,
  title={Repetitive readout of a single electronic spin via quantum logic with nuclear spin ancillae},
  author={Jiang, Liang and Hodges, JS and Maze, JR and Maurer, Peter and Taylor, JM and Cory, DG and Hemmer, PR and Walsworth, Ronald Lee and Yacoby, Amir and Zibrov, Alexander S and others},
  journal={Science},
  volume={326},
  number={5950},
  pages={267--272},
  year={2009},
  publisher={American Association for the Advancement of Science}
}

@article{waldherr2014quantum,
  title={Quantum error correction in a solid-state hybrid spin register},
  author={Waldherr, Gerald and Wang, Yiqing and Zaiser, S and Jamali, M and Schulte-Herbr{\"u}ggen, T and Abe, H and Ohshima, T and Isoya, J and Du, JF and Neumann, P and others},
  journal={Nature},
  volume={506},
  number={7487},
  pages={204--207},
  year={2014},
  publisher={Nature Publishing Group UK London}
}

@article{nickerson2014freely,
  title={Freely scalable quantum technologies using cells of 5-to-50 qubits with very lossy and noisy photonic links},
  author={Nickerson, Naomi H and Fitzsimons, Joseph F and Benjamin, Simon C},
  journal={Physical Review X},
  volume={4},
  number={4},
  pages={041041},
  year={2014},
  publisher={APS}
}

@article{fowler2012surface,
  title={Surface codes: Towards practical large-scale quantum computation},
  author={Fowler, Austin G and Mariantoni, Matteo and Martinis, John M and Cleland, Andrew N},
  journal={Physical Review A—Atomic, Molecular, and Optical Physics},
  volume={86},
  number={3},
  pages={032324},
  year={2012},
  publisher={APS}
}

@article{dur1999quantum,
  title={Quantum repeaters based on entanglement purification},
  author={D{\"u}r, Wolfgang and Briegel, H-J and Cirac, Juan Ignacio and Zoller, Peter},
  journal={Physical Review A},
  volume={59},
  number={1},
  pages={169},
  year={1999},
  publisher={APS}
}

@article{knill2005quantum,
  title={Quantum computing with realistically noisy devices},
  author={Knill, Emanuel},
  journal={Nature},
  volume={434},
  number={7029},
  pages={39--44},
  year={2005},
  publisher={Nature Publishing Group UK London}
}

@article{
stolk24metropolitan,
author = {Arian J. Stolk  and Kian L. van der Enden  and Marie-Christine Slater  and Ingmar te Raa-Derckx  and Pieter Botma  and Joris van Rantwijk  and J. J. Benjamin Biemond  and Ronald A. J. Hagen  and Rodolf W. Herfst  and Wouter D. Koek  and Adrianus J. H. Meskers  and René Vollmer  and Erwin J. van Zwet  and Matthew Markham  and Andrew M. Edmonds  and J. Fabian Geus  and Florian Elsen  and Bernd Jungbluth  and Constantin Haefner  and Christoph Tresp  and Jürgen Stuhler  and Stephan Ritter  and Ronald Hanson },
title = {Metropolitan-scale heralded entanglement of solid-state qubits},
journal = {Science Advances},
volume = {10},
number = {44},
pages = {eadp6442},
year = {2024},
doi = {10.1126/sciadv.adp6442},
URL = {https://www.science.org/doi/abs/10.1126/sciadv.adp6442},
eprint = {https://www.science.org/doi/pdf/10.1126/sciadv.adp6442}}

@article{azuma2023quantum,
  title={Quantum repeaters: From quantum networks to the quantum internet},
  author={Azuma, Koji and Economou, Sophia E and Elkouss, David and Hilaire, Paul and Jiang, Liang and Lo, Hoi-Kwong and Tzitrin, Ilan},
  journal={Reviews of Modern Physics},
  volume={95},
  number={4},
  pages={045006},
  year={2023},
  publisher={APS}
}

@article{metsch2019initialization,
  title={Initialization and readout of nuclear spins via a negatively charged silicon-vacancy center in diamond},
  author={Metsch, Mathias H and Senkalla, Katharina and Tratzmiller, Benedikt and Scheuer, Jochen and Kern, Michael and Achard, Jocelyn and Tallaire, Alexandre and Plenio, Martin B and Siyushev, Petr and Jelezko, Fedor},
  journal={Physical review letters},
  volume={122},
  number={19},
  pages={190503},
  year={2019},
  publisher={APS}
}

@article{harris2023hyperfine,
  title={Hyperfine spectroscopy of isotopically engineered group-IV color centers in diamond},
  author={Harris, Isaac BW and Michaels, Cathryn P and Chen, Kevin C and Parker, Ryan A and Titze, Michael and Arjona Mart{\'\i}nez, Jes{\'u}s and Sutula, Madison and Christen, Ian R and Stramma, Alexander M and Roth, William and others},
  journal={PRX Quantum},
  volume={4},
  number={4},
  pages={040301},
  year={2023},
  publisher={APS}
}

@article{marcomini2026,
  title={Gate Optimization via Efficient Two-Qubit Benchmarking for NV Centers in Diamond},
  author={Marcomini, Alessandro and Vetter, Philipp J and Calarco, Tommaso and Motzoi, Felix and Jelezko, Fedor and M{\"u}ller, Matthias M},
  journal={arXiv preprint arXiv:2603.08581},
  year={2026}
}
\end{document}